\documentclass{article}

\usepackage{PRIMEarxiv}

\usepackage[utf8]{inputenc} 
\usepackage[T1]{fontenc}    
\usepackage{hyperref}       
\usepackage{url}            
\usepackage{booktabs}       
\usepackage{amsfonts}       
\usepackage{nicefrac}       
\usepackage{microtype}      
\usepackage{lipsum}
\usepackage{fancyhdr}       
\usepackage{graphicx}       
\graphicspath{{media/}}     
\usepackage{setspace}
\usepackage{indentfirst}
 \usepackage{amsmath}
 \usepackage{booktabs}
 \usepackage{float} 
 \usepackage{multirow} 
 \usepackage[title]{appendix}
\title{Estimation of railway vehicle response for track geometry evaluation using branch Fourier neural operator
}

\author{
  Qingjing Wang \\
  qjwang@my.swjtu.edu.cn \\
   \And
  Wenhao Ding \\
  whding@my.swjtu.edu.cn\\
  \And
    Qing He \\
  qhe@swjtu.edu.cn\\
 \And
    Ping Wang \\
  wping@home.swjtu.edu.cn \\
}

\begin{document}
\maketitle

\begin{abstract}
In railway transportation, the evaluation of track geometry is an indispensable requirement to ensure the safety and comfort of railway vehicles. A promising approach is to directly use vehicle dynamic responses to assess the impact of track geometry defects. However, the computational cost of obtaining the dynamic response of the vehicle body using dynamics simulation methods is large. Thus, it is important to obtain the dynamic response of the vehicle–track coupled system efficiently and accurately. In this work, a branch Fourier neural operator (BFNO) model is proposed to obtain the dynamic response of the vehicle–track coupled system. The model takes into account the nonlinear relationship of the vehicle-track coupled system and realizes the fast and accurate estimation of the system dynamic response. The relative $L_2$ loss ($rLSE$) of BFNO model is 2.04\%, which is reduced by 64\%, compared with the traditional neural network (CNN-GRU). In the frequency domain, BFNO model achieves the effective estimation of the dynamic response of the system within the primary frequency range. Compared with the existing methods, our proposed model can make predictions at unseen time steps, enabling predictions from low to high time resolutions. Meanwhile, our proposed model is superior to commercial software in terms of efficiency. In the evaluation of track geometry, users can use pre-trained BFNO to obtain the dynamic response with almost no computational cost.
\end{abstract}


\section{Introduction}
\noindent 
The evaluation of track geometry is an indispensable requirement to ensure the safety and comfort of railway vehicles. Currently, the amplitude of measured track geometry parameters is used to detect and categorize track geometry defects. Although experience generally confirms the safety of this approach, the correlation between defect amplitude and vehicle response is relatively low. The performance of the bogies has a significant impact on the safety and reliability of high-speed trains \cite{ref1}. A promising approach is to directly use vehicle dynamic responses to assess the impact of track geometry defects \cite{ref2}. It is impractical to systematically measure all responses of the vehicle system,leading to the natural consideration of obtaining system responses through dynamic simulation methods. Although dynamic simulations have been widely applied in the railway industry, the limited application scope, incomplete functionality, and low computational efficiency of software solvers often necessitate the combined use of multiple software tools to complete simulation analysis \cite{ref3}. For instance, in Ref. \cite{ref4}, three commercial software (MATLAB, SIMPACK and ANSYS) were used in wheel polygonal wear simulation, which not only requires a high level of expertise of engineers but also results in high computational costs. Thus, it is important to obtain the dynamic response of the vehicle–track coupled system efficiently and accurately.
\par 
With the rise of big data mining techniques and artificial intelligence, some areas related to economy and personal safety, such as stock price forecasts \cite{ref5,ref6}, slope and flood disaster prediction \cite{ref7,ref8,ref9}, and structural optimisation \cite{ref10}, have developed rapidly. A number of machine learning approaches have also been proposed to provide faster numerical alternatives. Existing machine learning based methods can be broadly categorized into the following two categories: (1) Data-driven finite-dimensional operators for learning Euclidean space mappings from numerical simulation data \cite{ref11,ref12,ref13}, and (2) Physics-informed machine learning, which integrates data and physical models and implements them via neural networks or other kernel-based regression networks \cite{ref14,ref15}. The first approach, commonly used network architectures such as multilayer perceptron (MLP) \cite{ref16}, convolutional neural networks (CNN) \cite{ref17}, recurrent neural networks (RNN) \cite{ref18}, etc. Some data-driven models have also been used to obtain vehicle system responses. Kraft et al \cite{ref2} used a black-box model based on RNN neural networks to predict acceleration using track irregularity. Ye et al. \cite{ref19} proposed the MBSNet model, which uses short-term simulation results from a vehicle–track vertical coupled dynamics model to predict long-term system dynamic responses. However, data-driven models are prone to overfitting and therefore require large data sets. In addition, the results produced by these models are related to the specific grids of inputs and outputs used in the numerical simulation dataset \cite{ref20}. A representative example reflecting the second approach is physics-informed neural networks (PINNs), which use neural finite difference methods. The training process is often sensitive to selection of hyperparameters, retraining and fine-tuning of the PINN are required when the system parameters change.
\par
To address the above limitations, the concept of neural operators has been proposed. Neural operators directly learn infinite-dimensional mappings from any function parameter dependencies to solutions, with a key design choice being the selection of operator layers. Commonly used operator layers include Graph Neural Operator (GNO) \cite{ref21}, Low-rank Neural Operator (LNO) \cite{ref22}, Multipole Graph Neural Operator (MGNO) \cite{ref23}, and Fourier Neural Operator (FNO) \cite{ref24}. FNO utilizes the calculus properties of the Fast Fourier Transform for convolution operations, which provides the best performance compared to other operator layers \cite{ref22}. The FNO-based has shown excellent performance in multibody dynamics problems with great generalization ability \cite{ref25,ref26}. At the same time, the proposed DeepONet \cite{ref27} shares the same concept with the neural operator. The DeepONet processes data on two parallel networks (branch and stem networks), dealing with input and output functions, respectively. Then the mapping operators for parameterized partial differential equations can be learned by combing the outputs of the two networks. This model has derived a few applied research directions as well, including the inference of the electroconvection Multiphysics fields \cite{ref28} and reliability analysis of dynamic systems \cite{ref29}.
\par
In this paper, we extend the FNO-based architecture to vehicle–track spatially coupled dynamics systems, enabling the rapid acquisition of responses from the vehicle system. Considering the different frequency ranges of the car body and bogie frames, we propose the BFNO structure. Compared with CNN \cite{ref17}, CNN-GRU and original FNO model architectures, our proposed model has better estimation performance. 
\par
This study provides the following contributions:
\par
(1)	A vehicle–track spatially coupled dynamic model was built that considers the dynamic coupling between the wheel and rail.
\par
(2)	For the first time, we propose a branch Fourier neural operator (BFNO) architecture that separates the car body from bogie, enabling the independent learning of their respective mapping relationships within the system.
\par
(3)	Our proposed model is able to achieve good estimation in both time and frequency domains, and model achieves predictions from low to high time resolutions.
\par
The remainder of this paper is organized as follows. Section 2 introduces the vehicle–track spatially coupled dynamics model. Section 3 presents an overview of the proposed methodology, including FNO theory, the architecture of the BFNO model, equation weight normalization technique, and loss function design. Section 4 presents the generation of data, the selection of hyperparameters in neural networks and shows the comparison of estimation results between different models. Finally, conclusions and discussions are given in Section 5. 

\section{Vehicle–track spatially coupled dynamics model}
\begin{figure}[H]
    \centering 
    \includegraphics[width=0.7\textwidth]{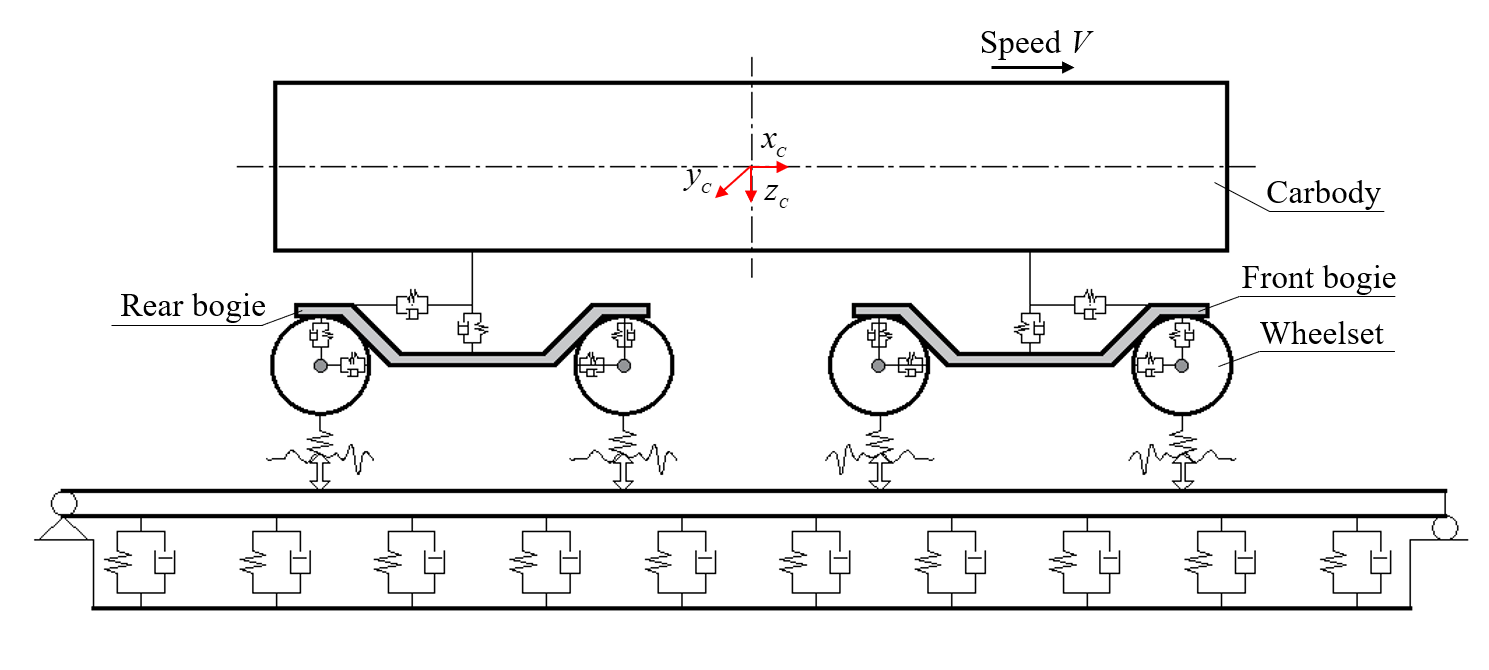} 
    \caption{Vehicle–track spatially coupled dynamics model.}
    \label{fig1} 
\end{figure}

\begin{figure}[H]
    \centering 
    \includegraphics[width=0.7\textwidth]{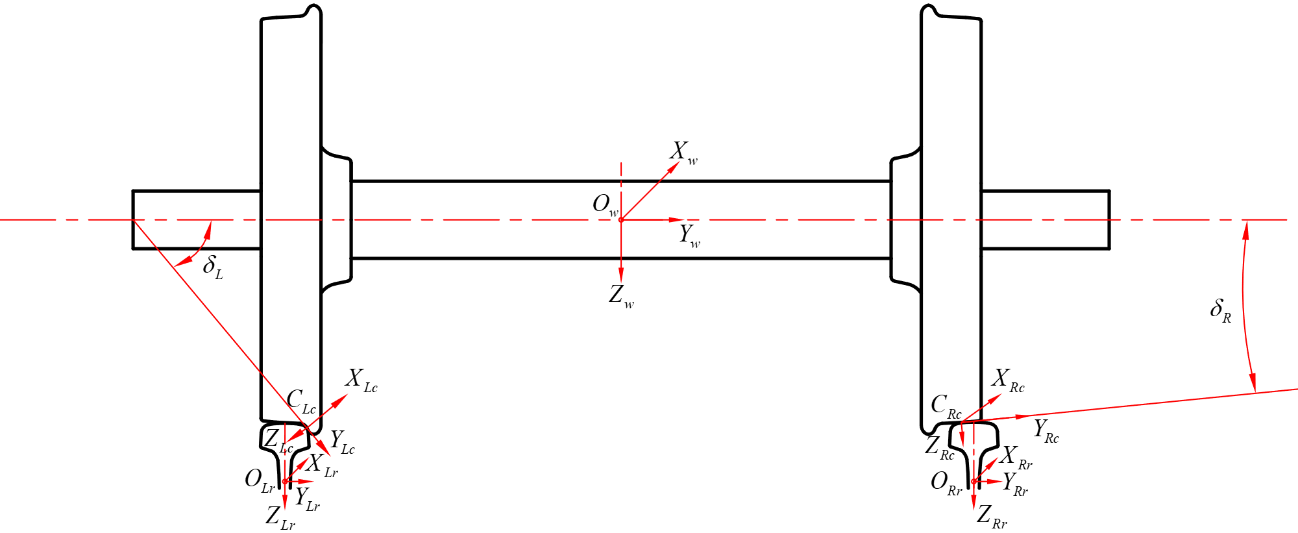} 
    \caption{Wheel–rail coupled model.}
    \label{fig2} 
\end{figure}

\noindent 
The vehicle–track spatially coupled dynamics model introduced in this study was proposed by Zhai \cite{ref30}. The schematic diagram is shown in Figure ~\ref{fig1}. The vehicle was set as a multi-rigid body assembly, including one car body, two bogie frames, and four wheelsets. The car body, bogie frames, and wheelsets were modelled as rigid bodies. Each rigid body was assigned five degrees of freedom (DOFs) including the lateral displacement Y, vertical displacement Z, roll angle $\phi$, yaw angle $\psi$, and pitch angle $\beta$. The longitudinal motion was known and characterised by a constant speed $V$. Thus, the total number of DOFs of the vehicle model was 35. The double-layer flexible track model is composed of two parallel steel rails, installed on sleepers through elastic fasteners. The first layer of the track is composed of rails, discretely supported on fasteners in the form of finite-length simply supported Euler beams. Both vertical and lateral bending and torsional deformation of the rails are taken into account. Each sleeper in the second layer of the track is treated as a rigid body with three degrees of freedom: lateral displacement, vertical displacement and roll. The connection between the rails and the sleepers (fasteners) and between the sleepers and the subgrade are represented by a set of linear spring-damping elements in the lateral and vertical directions. The vehicle and track subsystems were connected through wheel–rail interaction forces. To spatially couple the vehicle and track subsystems at the wheel–rail interface, we used the wheel–rail coupling model developed by Chen and Zhai \cite{ref31}, which considers the motions of the rail in the vertical, lateral, and torsional directions, as shown in Figure~\ref{fig2}. The wheel–rail interaction forces mainly include the wheel–rail normal contact and tangential forces. They were calculated using nonlinear Hertzian elastic contact and a combination of Kalker’s linear creep theory and the Shen-Hedrick-Elkins nonlinear model. For brevity, the detailed wheel–rail calculation method presented in \cite{ref32,ref33} is not presented here. 
\par
Vehicle–track spatially coupled dynamics is a typical multibody dynamics (MBD) problem. The vibration equation for the vehicle and track subsystems is expressed as Equation (1).
\begin{equation}
\mathbf{M}\ddot{\mathbf{X}}(t)+\mathbf{C}\dot{\mathbf{X}}(t)+\mathbf{K}\mathbf{X}(t)=\mathbf{P}(t)
\end{equation}
\par
\noindent 
where M, C, and K are the generalized mass, damping, and stiffness matrices, respectively; $\mathbf{X}$, $\dot{\mathbf{X}}$, and $\ddot{\mathbf{X}}$ are the vectors of displacement, velocity, and acceleration, respectively; P is the vector of the applied loads of the system; \textit{t} represents time.

\par
Inspired by the Newmark-$\beta$ implicit method, the Zhai method was used to obtain approximate solutions to the equations, as shown in Equation (2):
\begin{equation}
\left\{\begin{array}{l}
\mathbf{X}_{n+1}=\mathbf{X}_{n}+\dot{\mathbf{X}}_{n} \Delta t+\left(\frac{1}{2}+\psi\right) \ddot{\mathbf{X}}_{n} \Delta t^{2}-\psi \ddot{\mathbf{X}}_{n-1} \Delta t^{2} \\
\dot{\mathbf{X}}_{n+1}=\dot{\mathbf{X}}_{n}+(1+\varphi) \ddot{\mathbf{X}}_{n} \Delta t-\varphi \ddot{\mathbf{X}}_{n-1} \Delta t
\end{array}\right.
\end{equation}
\par
\noindent
where  \textit{t} is the time step, and $\psi$, $\varphi$  are free parameters that control the stability and numerical dissipation of the algorithm. The best calculation accuracy and numerical stability were achieved when the values of both parameters were 1/2. Subscripts \textit{n}-1, \textit{n}, and \textit{n}+1 represent the previous, current, and subsequent integration step, respectively.
\par
Equation (1) is a set of second-order ODEs. The vibration equations of the vehicle subsystem, including the car body, bogie frames, and wheelsets are coupled ordinary differential equations; however, the vibration equations of the track subsystem are partial differential equations (PDEs). To solve Equation (1), they must be converted into ODEs.
\par
In engineering, the mode superposition method (MSUP) \cite{ref34} is often used to convert PDEs into ODEs. The MSUP is a classic and powerful technique that approximates the dynamic response of a structure by superposition of its small number of eigenmodes, as shown in Equation (3).
\begin{equation}
\boldsymbol{u}=\sum_{i=1}^{n} \phi_{i} q_{i}
\end{equation}
\par
\noindent
where $\boldsymbol{u}$ is the solution field (displacement); $\phi_{i}$ is the \textit{i}-th order mode shape function; $q_{i}$ is the coordinate value of the \textit{i}-th mode of $\boldsymbol{u}$; \textit{n} is the number of modes used.
\par
In this study, the rails were modelled as Euler beams, their vibration equations were fourth-order PDEs. The vertical vibration differential equation (Equation (4)) of the rail is taken as an example to show the process of PDE conversion to ODEs. The analytical  can be obtained by analytical derivation of simple mechanical structures. The analytical mode-shape functions for the rails are described in Equation (5). The solution of Equation (4) can be expressed using Equation (6).
\begin{equation}
E I_{Y} \frac{\partial Z_{r}(x, t)}{\partial x^{4}}+m_{r} \frac{\partial^{2} Z_{r}(x, t)}{\partial t^{2}}=-\sum_{i=1}^{N} F_{V i}(t) \delta\left(x-x_{i}\right)+\sum_{j=1}^{4} P_{j} \delta\left(x-x_{w j}\right)
\end{equation}
\begin{equation}
Z_{k}(x)=\sqrt{\frac{2}{m_{r} l}} \sin \frac{k \pi x}{l}
\end{equation}
\begin{equation}
Z_{r}(x, t)=\sum_{k=1}^{N_{V}} Z_{k}(x) q_{VK}(t)
\end{equation}
\par
\noindent
where $E I_{Y}$ is the rail bending stiffness to the y-axle; $Z_{r}(x, t)$ is the vertical displacement o f the rail; $F_{V i}$ is the vertical dynamic forces at the \textit{i}-th rail-supporting point; $\delta(x)$is the Dirac delta function; $x_{w j}$is the coordinate of the \textit{j}-th wheel, and $N$ is the number of sleepers under the rail; $N_{V}$ is the highest order of the vertical mode shape of the rail.
\par
For the intercepted modal order $N_{\mathrm{V}}$, it is necessary to reach the frequency and then more than twice the effective frequency of the analyzed rail. Substituting Equation (6) into Equation (4), multiplying both sides by $Z_{h}(x)\left(h=1,2, \cdots, N_{V}\right)$, integrating over [0, \textit{l}], and using the orthogonality between the different modes, we obtain
\begin{equation}
\ddot{q}_{\mathrm{V} k}(t)+\frac{E I_{Y}}{m_{\mathrm{r}}}\left(\frac{k \pi}{l}\right)^{4} q_{\mathrm{V} k}(t)=-\sum_{i=1}^{N} F_{\mathrm{V} i} Z_{k}\left(x_{\mathrm{s} i}\right)+\sum_{j=1}^{4} P_{j} Z_{k}\left(x_{\mathrm{w} j}\right)\left(k=1-N_{\mathrm{V}}\right)
\end{equation}
\par
Equation (4) was transformed from fourth-order PDEs into second-order ODEs, each of which describes the motion of one mode of the continuum. Thus far, all equations in the vehicle–track coupled system are ODEs, collectively represented as Equation (1).

\section{Methodology}
\noindent
The PDEs of rail vibration were transformed into ODEs. These ODEs, together with those describing the motion of the vehicle system, form a system of coupled differential equations, described as
\begin{equation}
\mathcal{M}(p, f, u, t)=0, \in D
\end{equation}
\par
\noindent
where \textit{p} and \textit{f} denote the physical parameters and system excitation parameters with $n_{p}$ and $n_{f}$ respectively; \textit{u} represents the solutions for all quantities in $\mathcal{M}$, and $D$ is the bounded time domain. 
\par
Our goal is to learn a mapping between two infinite-dimensional spaces from a finite set of input parameters and output parameters by constructing a neural network to approximate the solution operator of Equation (8). Let $a=(p, f) \in \mathcal{A} \subset \mathbb{R}^{n_{p}+n_{f}}$ be the variable parameter set for the vehicle–track system, and $u \in \mathcal{U}$ be the solution set with $n_{dof}$ elements. We aim to build an approximation $\mathcal{G}_{\theta}: A \times \theta \rightarrow \mathcal{U}$ of the solution operator for the coupled differential equation group $\mathcal{M}$. We demonstrate in this section that the proposed BFNO architecture learns $\mathcal{G}_{\theta}$. Figure~\ref{fig4} provides a schematic of the BFNO architecture. 
\subsection{Fourier neural operator}
\noindent
The FNO proposed by Li et al. \cite{ref24} was formulated as a generalization of standard deep neural networks to operator settings. A neural operator learns a mapping between two infinite dimensional spaces from a finite collection of observed input-output pairs. 

\paragraph{Definition 1 (Neural operator $\mathcal{G}_{\theta}$) Define the neural operator}
\begin{equation}
\mathcal{G}_{\theta}:=Q \circ \sigma\left(W_{L}+\mathcal{K}_{L}\right) \circ \cdots \circ \sigma\left(W_{1}+\mathcal{K}_{1}\right) \circ P
\end{equation}
\par
\noindent
where $P$ and $Q$ are pointwise neural networks that lift the input $a \in \mathcal{A}$ to a higher-dimension channel space and output $u \in \mathcal{U}$ by projecting it back to the target dimension.
\par
The model stack has L layers and the iterative update from layer   to layer ($v_{j} \mapsto v_{j+1}$) in the model is implemented by $\sigma\left(W_{j}+\mathcal{K}_{j}\right)$ ; $W_{j}$ are pointwise linear operators (matrices); $\mathcal{K}_{j}$ are integral kernel operators, and $\sigma$ are fixed activation functions. 
\paragraph{Definition 2 ($\mathcal{K}$) Define the Fourier convolution operator}
\begin{equation}
\left(\mathcal{K} v_{l}\right)=\mathcal{F}^{-1}\left(R \cdot T_{K}\left(\mathcal{F} v_{l}\right)\right.
\end{equation}
\noindent
where $\mathcal{F}$ and $\mathcal{F}^{-1}$ denote the Fourier transform and its inverse transform, respectively; $R$ is the Fourier transform of a periodic function, and $T_{K}$ is a fixed truncation that restricts the input to the lowest \textit{K} Fourier modes.
\begin{figure}[H]
    \centering 
    \includegraphics[width=0.7\textwidth]{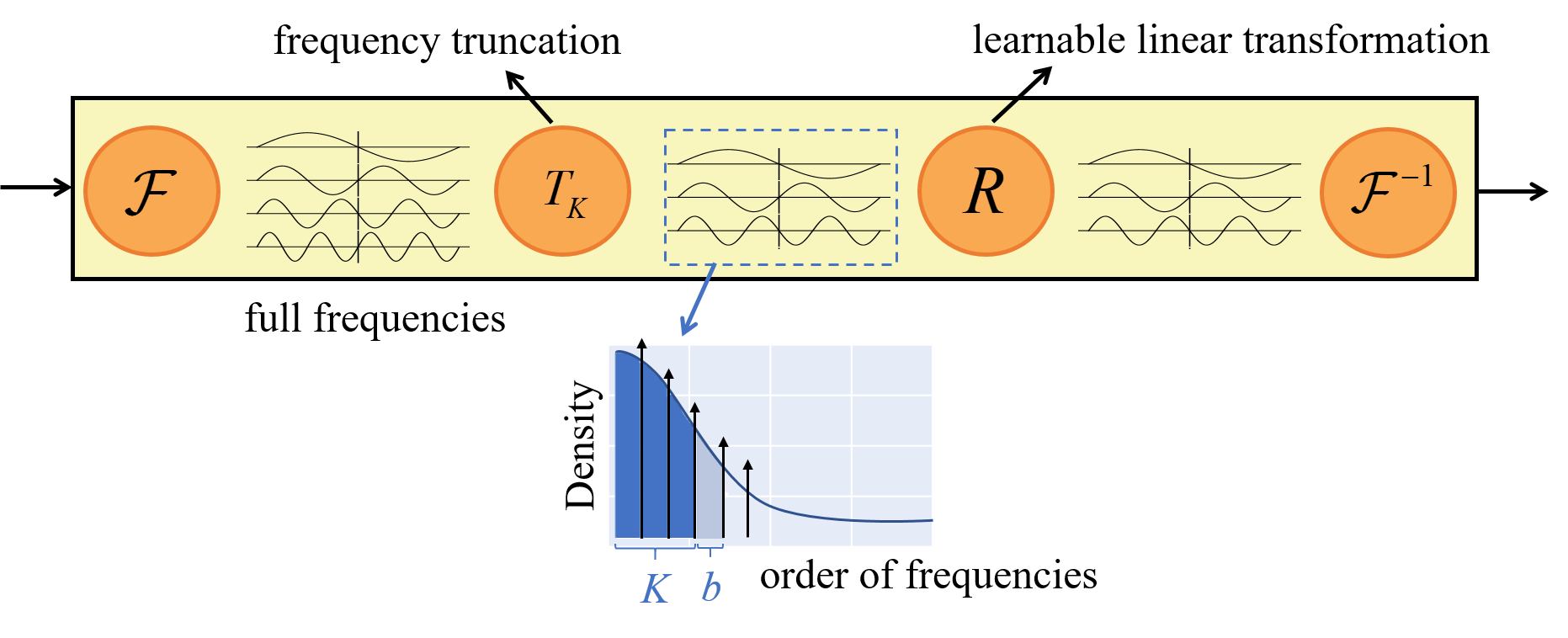} 
    \caption{Top: Fourier convolution operator in FNO. Apply a Fourier transform $\mathcal{F}$ and truncate the complete set of input frequencies to the K lowest frequencies using a fixed truncation $T_K$. Apply a learnable linear transform $R$ and apply the inverse Fourier transform $\mathcal{F}^{-1}$. Bottom: truncation of frequency modes. The solid areas indicate how much information in the spectrum is explained by the current mode $K$.}
    \label{fig3} 
\end{figure}
\subsection{BFNO architecture}
The FNO consists of a frequency-truncation $T_{K}$ in each layer that only allows the lowest \textit{K} Fourier modes to propagate the input information, as shown in Figure~\ref{fig3}. Although the frequency truncation $T_{K}$ ensures the discretization-invariance of the FNO, it is still challenging to select an appropriate number of effective modes \textit{K} because it is task-dependent and requires careful hyperparameter tuning. Setting \textit{K} too small results in too few frequency modes, with insufficient information for learning the solution operator, leading to underfitting. A large \textit{K} with too many effective frequency modes may encourage the FNO to insert noise into the high-frequency components. This results in overfitting and is computationally expensive \cite{ref35}.
\begin{figure}[H]
    \centering 
    \includegraphics[width=0.7\textwidth]{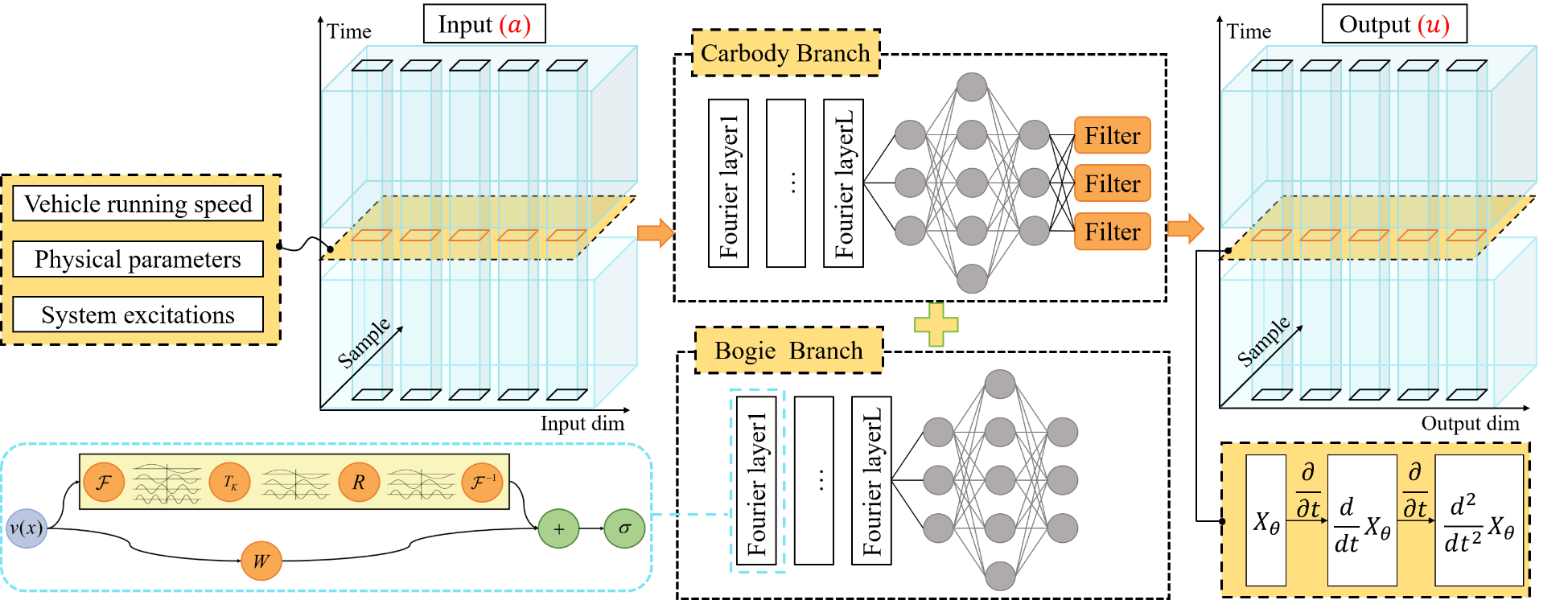} 
    \caption{Network architecture of BFNO.}
    \label{fig4} 
\end{figure}
\par
For the vehicle–track system, the frequency domain ranges corresponding to the car body, bogie frames, wheelsets, and rails are different. Setting the same K value may cause underfitting or overfitting. To address these concerns, we propose a BFNO architecture with two Fourier branches (Figure~\ref{fig4}). The BFNO architecture contains the following three steps:
\par
1. Lift the input $a \in \mathcal{A}$ to a higher-dimension channel space through a fully connected neural network transformation $P$.
\par
2. Apply iterative Fourier layers: $v_{1} \mapsto \ldots \mapsto v_{L}$ where $v_j$ for $j=1, \ldots, L$ , and project $v_j$ back to the original space using a fully connected neural network transformation $Q$. Specially, a low-pass filtering layer is added to the car body branch.
\par
3. Connecting the output structures of the car body branch and the bogie branch as the model output $u \in \mathcal{U}$.
\par
Fundamentally, the BFNO is used to extract the feature relationship between the input and output data in the frequency domain, a fully connected layer is used to combine the features and output approximate solutions. The neural operators only output solutions to Equation (8), and their derivatives were computed using a differentiation operation. 
\subsection{Equation weight normalization technique}
\noindent
The differential equation group $\mathcal{M}$ in a vehicle–track coupled system consists of multiple coupled ODEs. Considering the units and numerical differences in the different degrees of freedom, there are significant variations in the magnitude of the data corresponding to each equation. Use of normalization techniques reduces scale differences in the input data, improving the training effectiveness and performance of the neural networks. We propose an equation weight normalization technique that effectively leverages prior physical knowledge \cite{ref26}. The equation weight normalization technique assigns a corresponding weight factor $\lambda$ to each ODE in each dataset, ensuring that the gradient magnitudes across different ODEs are similar during the learning process and improving overall performance. The complete EN algorithm is presented as Algorithm 1.
\par
\begin{table}[htbp]  
\begin{center}  
\begin{tabular*}{\hsize}{@{}@{\extracolsep{\fill}}l@{}} 
        \toprule[1.5pt]  
        \textbf{Algorithm 1} Compute weight $\lambda_{ij}$ for $j^{th}$ the ODE in $i^{th}$ the dataset \\
        \midrule  
       \textbf{for} $i=1 \rightarrow N$ \textbf{do}\\
       \quad Extract data  set $\left\{a_{i}, u_{i}\right\}$\\
       \quad\textbf{for}  $j=1 \rightarrow n_{dof}$ \textbf{do}\\
         \quad\quad$u=u_{i}(:, j)$ \\
         \quad\quad$p_{0} \Leftarrow \varepsilon(r * \sigma(u), u), p_{1} \Leftarrow \varepsilon(r * \sigma(\dot{u}), \dot{u}), p_{2} \Leftarrow \varepsilon(r * \sigma(\ddot{u}), \ddot{u})$ \\
         \quad\quad$s_{0}(:, j) \Leftarrow u+p_{0}, s_{1}(:, j) \Leftarrow \frac{\partial u}{\partial t}+p_{1}, s_{2}(:, j) \Leftarrow \frac{\partial^{2} u}{\partial t^{2}}+p_{2}$ \\
        \quad\textbf{end for}\\
        \quad$L=\max ^{T}\left|\mathcal{M}\left(a_{i}, s_{0}, s_{1}, s_{2}\right)\right|$ \\
        \quad\textbf{for}  $j=1 \rightarrow n_{dof}$ \textbf{do} \\
         \quad\quad$\lambda_{i j}=\frac{r}{L(j)}$ \\
        \textbf{end for}\\
        \bottomrule  
    \end{tabular*}
    \label{tab1}
\end{center}
\end{table}
\par
\noindent
where $\sigma(x)$ represents the standard deviation of $x$, and $\sigma(a,b)$  generates a random sequence following a uniform distribution on the interval $(-a,a)$ with the same dimensions as $b$.  $\operatorname{Max}^{T}|\mathcal{M}|$computes the extremum of each ODE in   in the time domain. $r$ is the sensitivity factor, indicating an acceptable level of signal error. In this study, we set $r$ as 2\% as a greater penalty.

\subsection{Loss function design}
\noindent
In general, the model performs optimally when the same loss criterion is used for both training and testing. The choice of the loss function plays a crucial role. In this work, we use the relative $L_2$ loss ($rLSE$) to measure the performance in our model. The $L_2$ loss and its square, the mean squared loss (MSE), are common choices for testing criteria in deep learning. We observe that training the model using relative loss has beneficial effects in terms of normalization and regularization, preventing overfitting. The $L_2$ loss is applied to both the original output ($y(t)$), the first derivative of the output ($\partial y / \partial t$), and the second derivative of the output ($\partial^{2} y / \partial t^{2}$), is written as:
\begin{equation}
L(y, \hat{y})=w_{1} \frac{\|y-\hat{y}\|_{2}}{\|y\|_{2}}+w_{2} \frac{\|\partial y / \partial t-\partial \hat{y} / \partial t\|_{2}}{\|\partial y / \partial t\|_{2}}+w_{3} \frac{\left\|\partial^{2} y / \partial t^{2}-\partial^{2} \hat{y} / \partial t^{2}\right\|_{2}}{\left\|\partial^{2} y / \partial t^{2}\right\|_{2}}
\end{equation}
\par
\noindent
where $\hat{y}$ is the predicted output, $\partial \hat{y} / \partial t$ is the first derivative of the predicted output, $\partial^{2} \hat{y} / \partial t^{2}$ is the second derivative of the predicted output, and $w_{1}, w_{2}, w_{3}$ are the weights of each error. Use of derivative loss was motivated by the engineering community's primary interest in derivatives rather than solutions. For example, sound in nature is typically associated with vibration speed (first derivative), whereas human comfort is linked to acceleration (second derivative). 
\section{Results}
\noindent
This section compares 4 types of model architectures: CNN, CNN-GRU, FNO and BFNO. All models are trained on the proposed loss function. Detailed parameters for each model are summarized in Appendix B to Appendix E.
\subsection{Data preparation}
\begin{figure}[H]
    \centering 
    \includegraphics[width=0.7\textwidth]{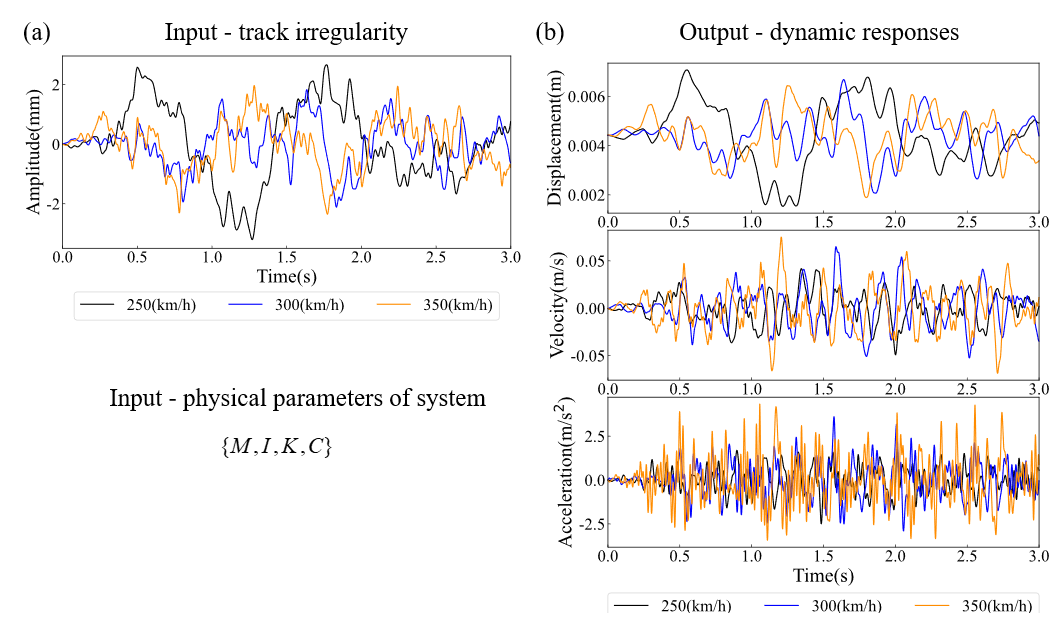} 
    \caption{Example of mapping between (a) input variable parameters to (b) output dynamic responses. (a) running speed, physical parameters of system and track irregularities. (b) vertical vibration of car body.}
    \label{fig5} 
\end{figure}

\noindent
The input and output parameters for the training, validation, and test data set were generated using the Zhai method, described in Section 2. We set the three types of input parameters in each numerical simulation case to be variable: running speed, physical parameters of system and system excitations. As shown in Figure~\ref{fig5}, the running speed is considered to vary from 250 to 350 km/h. Physical parameters including mass($M$), mass moment of inertia($I$), stiffness($K$) and damping($C$).The complete range of physical parameters is shown in Appendix A. The output parameters included the dynamic responses for the vehicle–track system.

\par
\begin{figure}
    \centering 
    \includegraphics[width=0.7\textwidth]{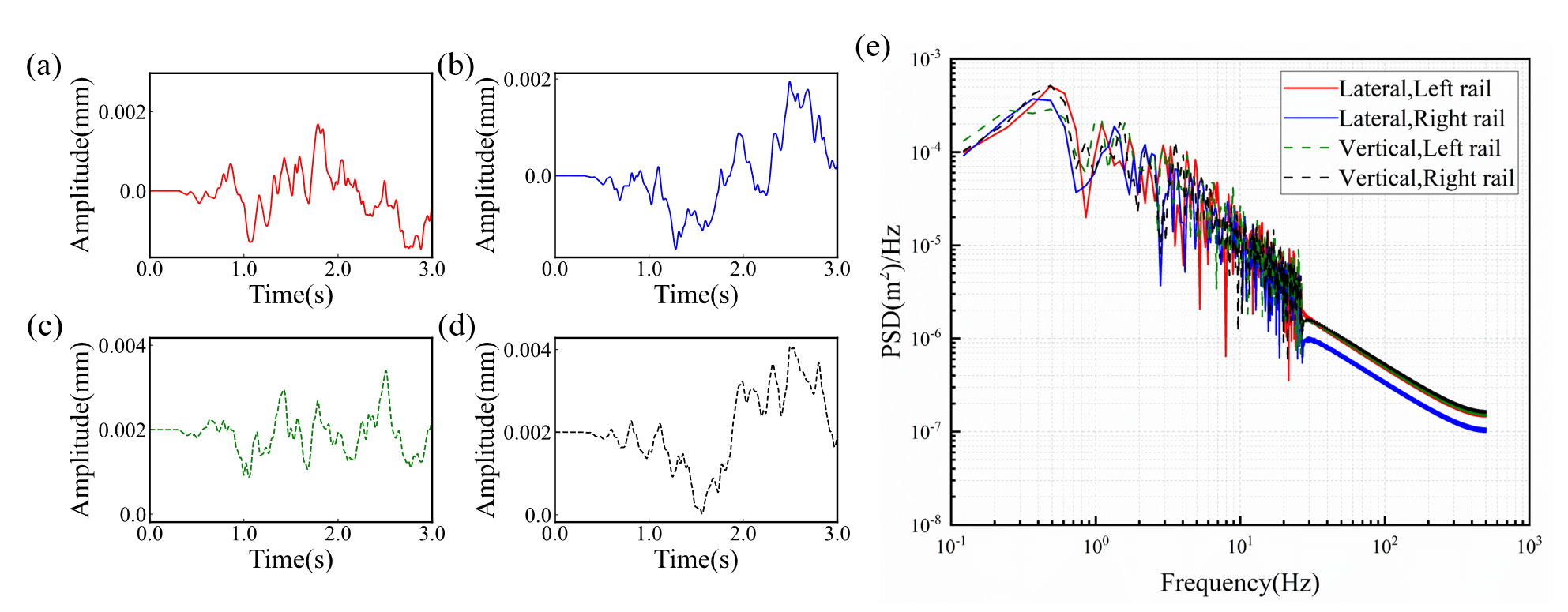} 
    \caption{Simulation samples of track irregularities: (a) left rail (lateral); (b) right rail (lateral); (c) left rail (vertical); (d) right rail (vertical); (e) PSDs.}
    \label{fig6} 
\end{figure}

In this study, only track irregularities were included in the system excitations. The track irregularities present on an actual line are superimposed by random irregularities of different wavelengths, phases, and assignments, and are complex stochastic processes related to line mileage. The power spectral density (PSD) is the most commonly used statistical function for representing random track irregularities. It is usually considered as a stationary stochastic process; the statistical characteristics of random track irregularities can only be obtained from field measurements. In engineering, PSD charts are often used to describe the relationship between spectral densities and corresponding frequencies. We used the classical spectral method \cite{ref30} to generate spatial domain simulation samples of track irregularities (Figure~\ref{fig6}) and used the Chinese high-speed ballastless track spectrum as a PSD function. The sampling frequency was set as 1000 Hz and the wavelength ranged from 1–100 m. As input parameters, the superimposed irregularities were used as track irregularities, and were composed of vertical and lateral irregularities. In total, 10,000 sets of training data, 1,000 sets of validation data, and 1,000 sets of test data were generated using a multibody dynamics simulation (MBS) based on the Zhai method. 
\subsection{Hyperparameter selection}
\noindent
Selection of hyperparameters is crucial in neural networks, the main hyperparameter tuning methods include manual tuning, grid search, random search, and Bayesian optimization. Bayesian optimization uses the performance of previously searched parameters to infer the next step, improving the search efficiency \cite{ref36}. To obtain the best performance of the BFNO model, we used Bayesian hyperparameter optimization to search for the minimum value, with the objective function as the minimum loss of the validation set. Some parameters in the neural network model are nested in the activation function they select. These parameters have default values that have little effect on model optimization; thus, they are generally not considered in Bayesian optimization. The main difference between the car body branch and the bogie branch in the BFNO architecture is the Fourier modes K. Therefore, during the process of Bayesian optimization, the hyperparameters of the Fourier layer and the fully connected layer are set to the same values.
\par
\begin{figure}[H]
    \centering 
    \includegraphics[width=0.7\textwidth]{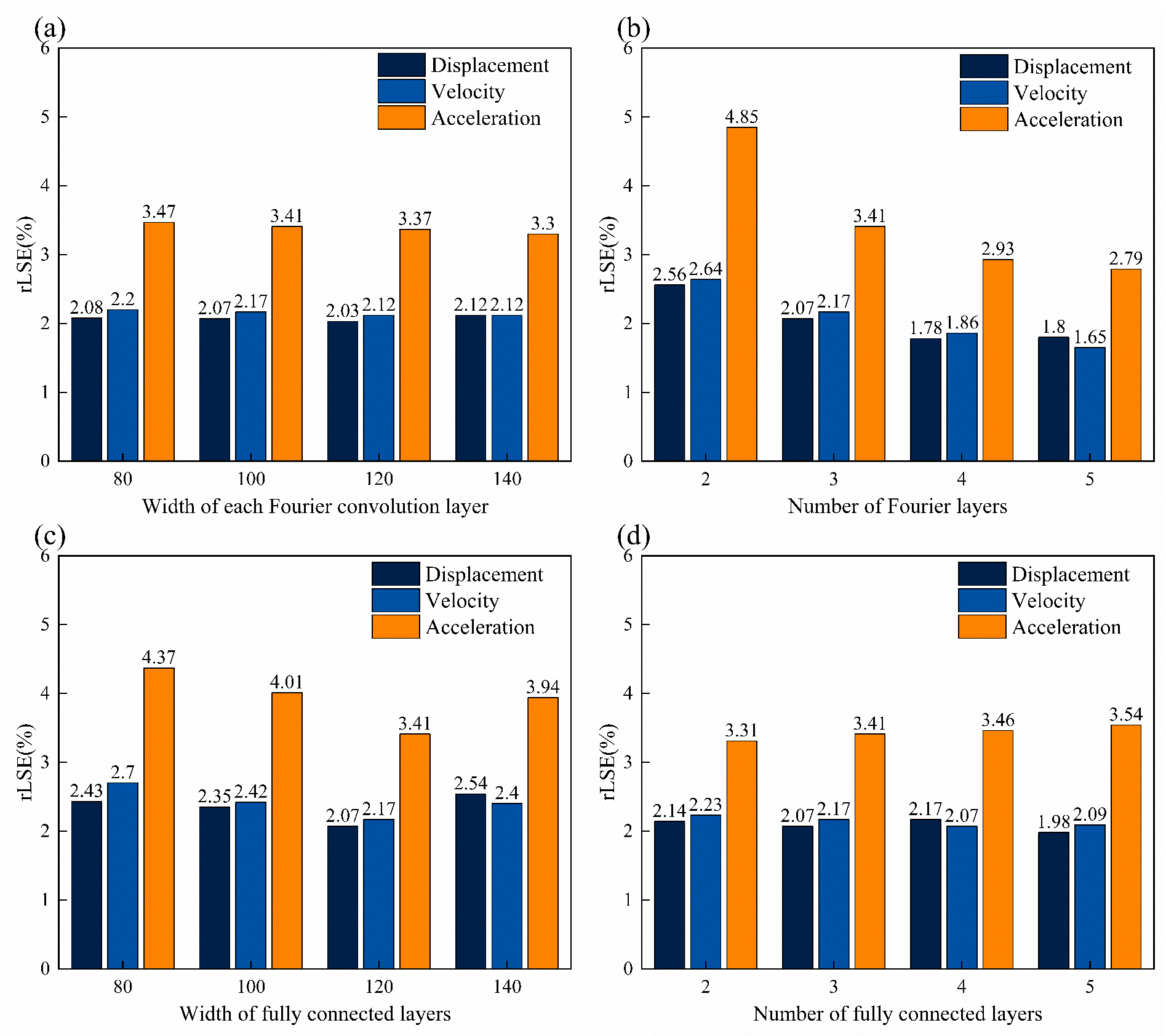} 
    \caption{Hyperparameter analysis for proposed model.}
    \label{fig7} 
\end{figure}

\begin{table}
\caption{Hyperparameter optimization and settings.}
\centering
{\begin{tabular}{llll} \toprule
 Type  & Hyperparameter & Range of selection & Optimal value\\ \midrule
\multirow{4}{2cm}{Automatic} & Number of Fourier layers & [3,6] & 4\\ 
    & Width of each Fourier convolution layer& [50,200]& 127\\
    &Number of fully connected layers&[3,6]&3\\
    &Width of fully connected layers&[50,200]&131\\
\midrule
 \multirow{4}{2cm}{Manual} & Fourier modes used in car body branch & /&100\\
    &Fourier modes used in bogie branch	&/	&500\\
    &Batch size	&/	&32\\
    &Learning rate	&/	&0.001\\
  \bottomrule
\end{tabular}}
\label{table1}
\end{table}

Figure~\ref{fig7} presents the hyperparameter analysis results for model. As can be seen from the figure, the model has limited sensitivity to hyperparameters. Additionally, it can be seen that increasing the number of Fourier layers has the most significant positive impact on accuracy improvement. In the process of hyperparameter analysis, there are 13 initial points, and the model is optimized 10 times considering the computational costs. Table 1 displays the optimized results of several critical hyperparameters whose broad ranges were determined from the results of manual tuning tests and other manually set hyperparameters. 
\par
As shown in Table~\ref{table1}, three Fourier layers and four fully connected layers were designed, the widths of each layer were set as 127 and 131, respectively. The Fourier mode used in car body branch was based on the frequency range of the car body response and was set as 100. The Fourier mode used in bogie branch was set according to the sampling frequency and was set as 500. Other parameters and techniques used in model included: (1) We found that model’s performance showed minimal sensitivity to the batch size. Therefore, the batch size was set to 32, the maximum value allowed by the GPU occupancy limit.; (2) Adam optimizer to update the local learning rate during the training process and enhance the robustness of the network; (3) the initial learning rate was set as 1e-3, the default value in the Adam optimizer, and was reduced to 75\% every 30 steps; (4) The ELU activation function (used in fully connected layers) has advantages with negative inputs and promotes training stability. 
\subsection{Performance analysis}
\noindent
\begin{figure}[H]
    \centering 
    \includegraphics[width=0.6\textwidth]{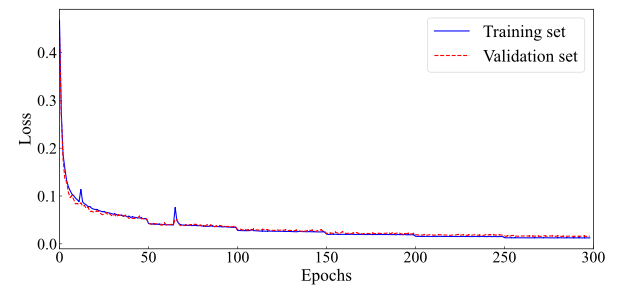} 
    \caption{The loss curves of the training set and the validation set.}
    \label{fig8} 
\end{figure}
The evaluation metrics we use were $rLSE$ (defined in Subsection 3.3) and Pearson correlation coefficient (PCC), its formula is shown in Equation (12).
\begin{equation}
P C C=\frac{\sum_{i=1}^{M}\left(\hat{y}_{i}-\overline{\hat{y}}\right)\left(y_{i}-\bar{y}\right)}{\sqrt{\sum_{i=1}^{M}\left(\hat{y}_{i}-\overline{\hat{y}}^{2}\right.} * \sqrt{\sum_{i=1}^{M}\left(y_{i}-\bar{y}\right)^{2}}},\left(\overline{\mathrm{y}}=\frac{1}{M} \sum_{i=1}^{M} y_{k}, \overline{\hat{y}}=\frac{1}{M} \sum_{i=1}^{M} \hat{y}_{k}\right)
\end{equation}
\par
\noindent
where $y_i$ and $\hat{y}_{i}$ are the \textit{i}-th actual and predicted data points, respectively, and $\bar{y}$ and $\overline{\hat{y}}$ are the expectations for the actual and predicted data points, respectively. The range of the PCC is from -1 to 1, a higher numerical value indicates a stronger positive correlation and better performance.
\par

\begin{figure}[H]
    \centering 
    \includegraphics[width=0.7\textwidth]{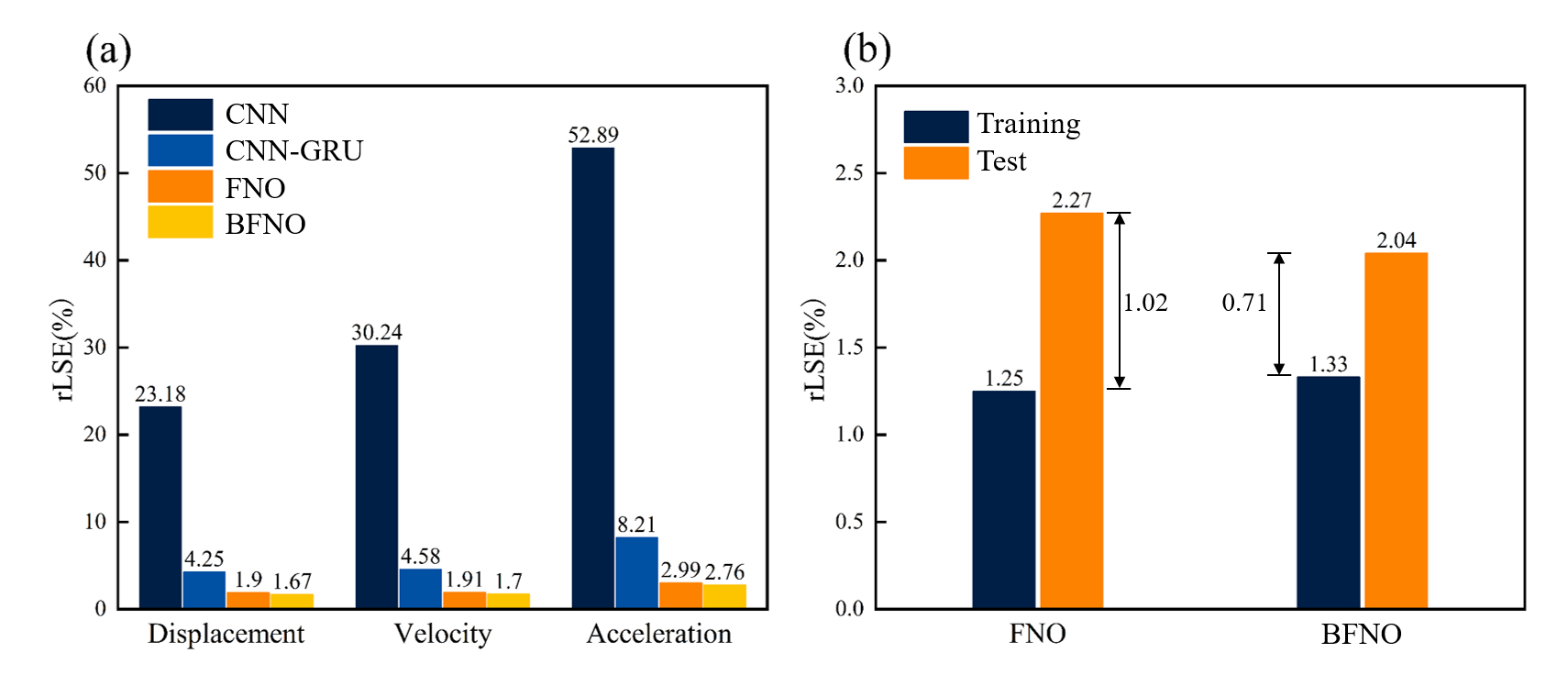} 
    \caption{Analysis of model performance: (a) Test error using CNN, CNN-GRU, FNO, and BFNO; (b) Average training and test error of FNO and BFNO.}
    \label{fig9} 
\end{figure}
\begin{figure}[H]
    \centering 
    \includegraphics[width=0.7\textwidth]{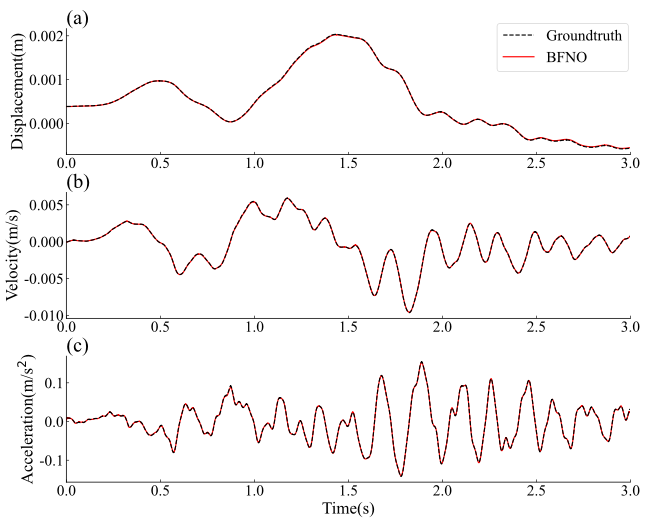} 
    \caption{Comparison of the vertical response between ground truth and model estimation results.}
    \label{fig10} 
\end{figure}
The training and the validation loss curves of the BFNO model are shown in Figure~\ref{fig8}. It can be seen that the loss curve remains stable when the epoch reaches 300 times. Figure~\ref{fig9}(a) demonstrates that the best performance for both the training and testing data set is achieved with the BFNO model. In contrast, the CNN model failed to capture the nonlinear features of the lateral vehicle dynamics, resulting in the poorest performance of the CNN model. Compared with CNN-GRU model, both the original FNO and BFNO consist of frequency-truncation $T_K$ in each layer that only allows the lowest $K$ Fourier modes to propagate the input information. Selecting an appropriate number of effective modes $K$ can filter out high frequency information from the input parameters. We can observe from Figure~\ref{fig9}(b) that the training errors of both structures were similar, but BFNO exhibited a larger generalization error (the difference between the test error and training error). The results of vertical dynamic response estimation using the BFNO method and MBS method are shown in Figure~\ref{fig10}. The displacement, velocity, and acceleration results obtained using the BFNO model had good accuracy. In the model training, only ground truth of displacement was provided, while velocity and acceleration were computed using numerical differentiation methods. In contrast to other studies that only provide ground truth of acceleration for the dynamic system, our approach in this paper in this paper more closely to physical principles. However, in the field of railway engineering, researchers pay more attention to acceleration information. Therefore, in the subsequent comparison of model performance, we only present the results related to acceleration. 
\par
\begin{figure}[H]
    \centering 
    \includegraphics[width=0.7\textwidth]{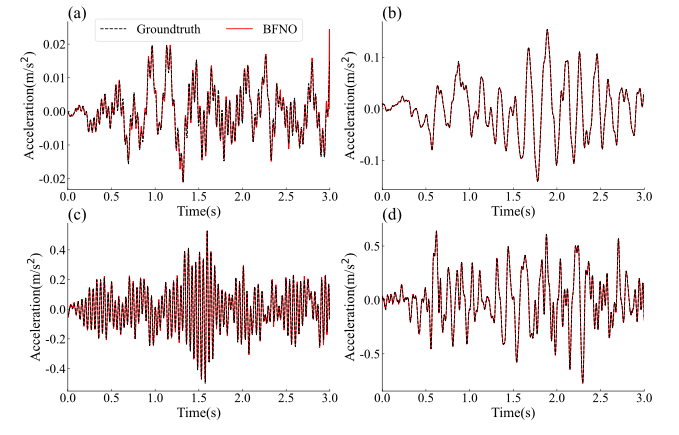} 
    \caption{The prediction results in the time domain. (a), (b) lateral and vertical acceleration of car body; (c), (d) lateral and vertical acceleration of bogie.}
    \label{fig11} 
\end{figure}
\begin{figure}
    \centering 
    \includegraphics[width=0.7\textwidth]{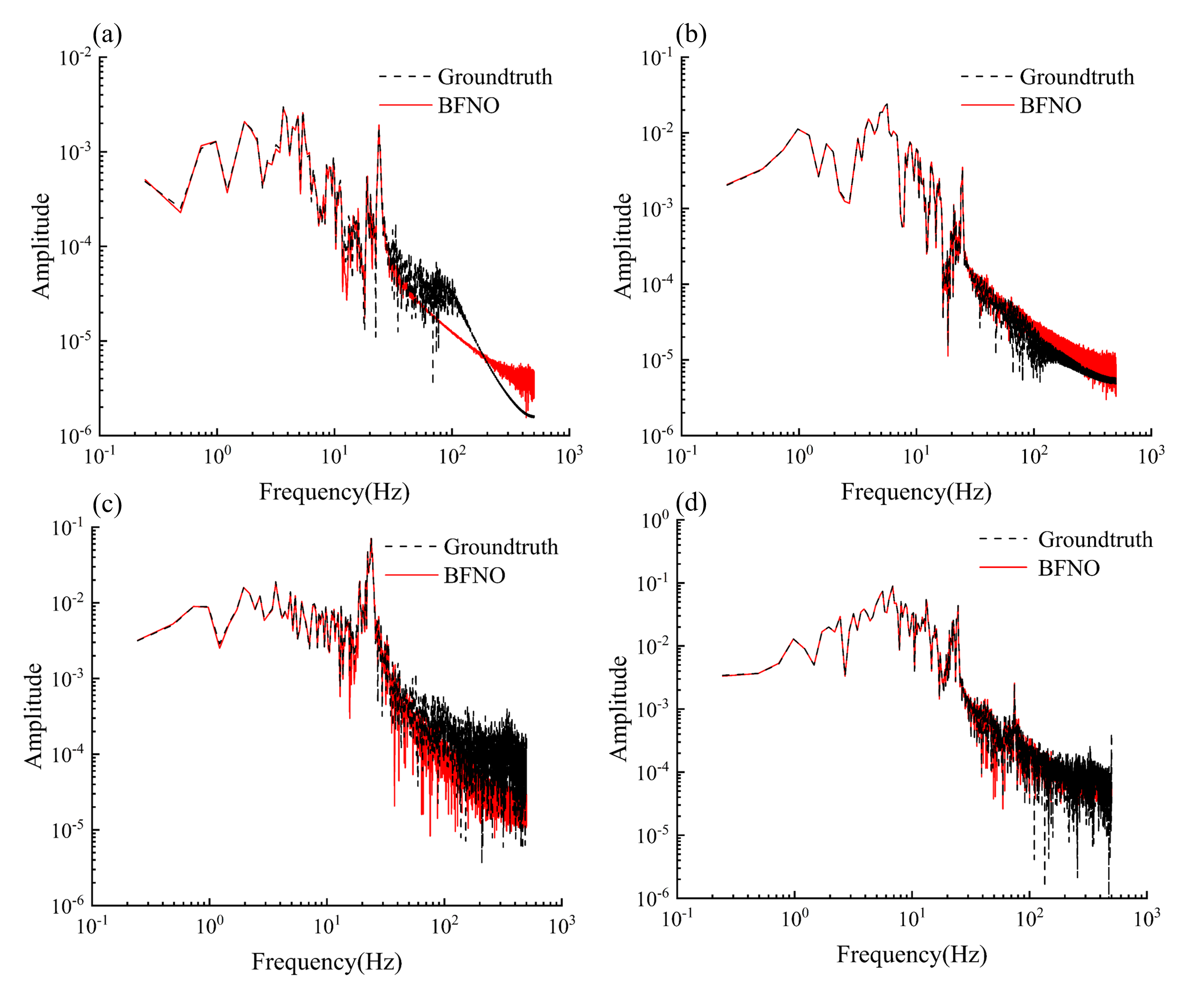} 
    \caption{The prediction results of frequency spectrum using the BFNO model.}
    \label{fig12} 
\end{figure}

In order to considering the average performance across the entire training and testing sets, the estimation results of BFNO and MBS methods are shown in Figure~\ref{fig11}. The BFNO has good performance in all of these parts, which can be quantified by low $rLSE$ loss and high PCC value (Table~\ref{table2}). Figure~\ref{fig12} shows the frequency spectrum curve inferred from the model prediction results. It can be seen that the estimated value and the ground truth are still very consistent within the primary frequency range.

\begin{table}[H]
\caption{Summary of Evaluation metric.}
\centering
{\begin{tabular}{llll} \toprule
 Part  & Acceleration& $rLSE$(\%) & PCC \\ \midrule
\multirow{2}{2cm}{Car body} & Lateral acceleration&4.16&0.9717\\
& Vertical acceleration&1.62&0.9836\\
\midrule
 \multirow{2}{2cm}{Bogie} & Lateral acceleration	&5.17&0.9220\\
 & Vertical acceleration&1.54&0.9864\\
  \bottomrule
\end{tabular}}
\label{table2}
\end{table}

\par
For the vehicle-track coupled dynamics problem addressed in this paper, the model is trained only once, and solving new cases with parameter variations no longer requires additional training. The trained machine learning models can be directly used for solving without further training. Selecting a 3-second segment of data for comparison, the MBS model computed in MATLAB required approximately 20 min; UM® and SIMPACK® required 214.73 s and 209.21 s, respectively. In contrast, the BFNO model running in in the Pytorch framework required only 3.69 s (on a device with an Intel Core i9-12900K processor and an Nvidia GeForce RTX 3060 Super graphic processing unit). Using the BFNO model is that its computing speed is significantly improved. 
\begin{figure}[H]
    \centering 
    \includegraphics[width=0.7\textwidth]{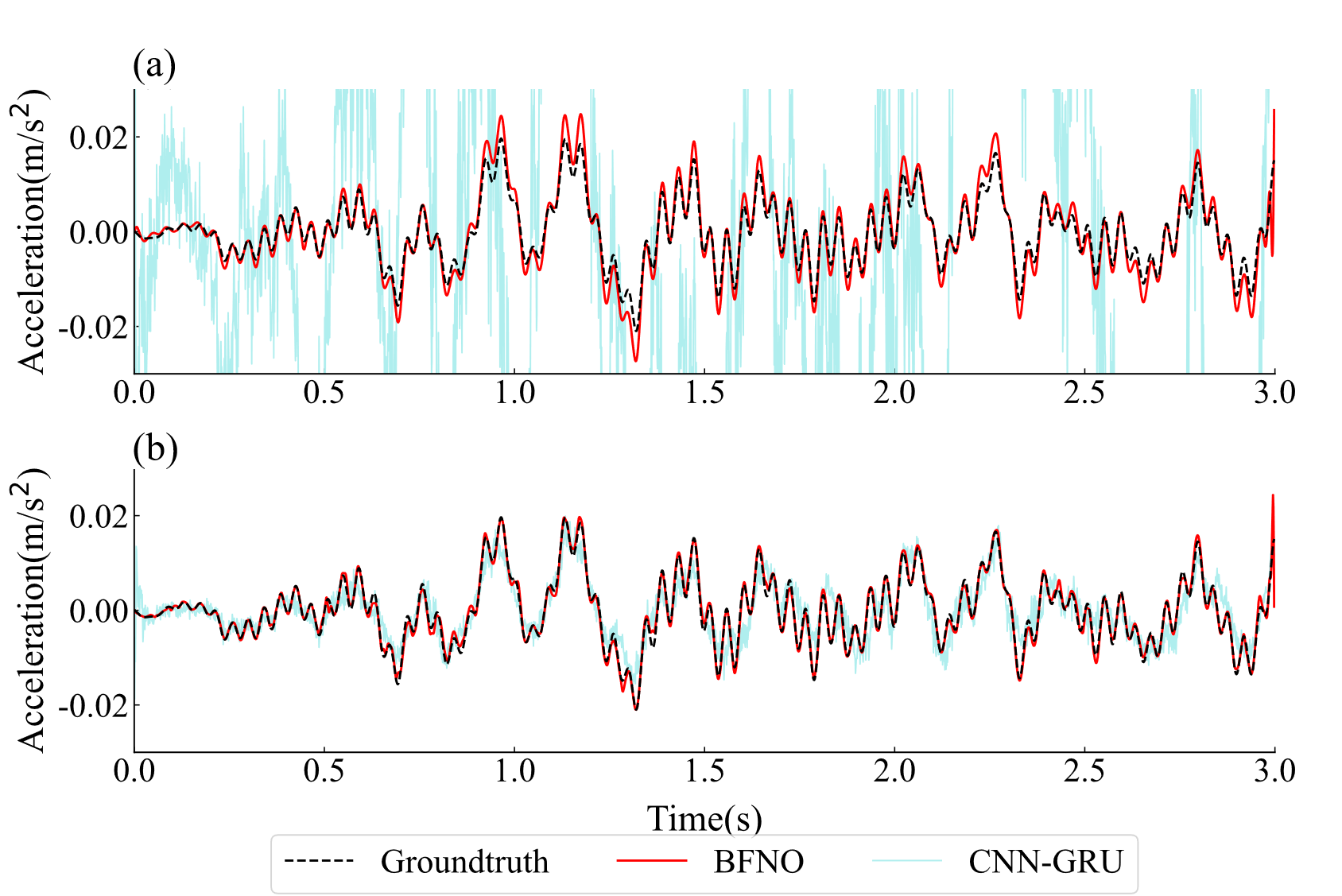} 
    \caption{Comparison of the lateral car body acceleration of BFNO and CNN-GRU in dataset A, B. (a) performance of dataset A, (b) performance of dataset B.}
    \label{fig13} 
\end{figure}
\subsection{Prediction at unseen time steps}
\noindent
FNO-based architectures are infinite-dimensional operators that can provide grid-invariant predictions. To demonstrate the performance of this technique, we test the CNN-GRU, BFNO at a temporal resolution not used during training, specifically, data at 0.003s time step for the training set and 0.001s time step for the test set. Table~\ref{table3} summarizes the evaluation metric of the new dataset (dataset A) compared to the 0.001s time step dataset for both training and testing (dataset B). In order to show the results clearly, we only compared the lateral car body acceleration under two datasets in Figure~\ref{fig13}.
\par
It can be observed from Figure~\ref{fig13}(a) that the CNN-GRU model exhibits poor estimation performance in dataset A, while the estimated values of the FNO model are almost the same as the ground truth. However, in some peaks and valleys, the estimated value of the FNO model is relatively larger than the ground truth. Compared to the dataset B, although the performance of the BFNO model decreases slightly with the change in the time-steps, the BFNO model still provides relatively good estimation at the unseen time without additional training. The model achieves predictions from low to high time resolutions. For larger multibody dynamic systems, the training set uses low time resolution data to reduce data generation time and memory usage. The test set uses high time resolution data to enhance the improve the prediction accuracy of the model, thereby meeting the requirements of both data utilization and precision.
\begin{table}[H]
\caption{Performance comparison of CNN-GRU, BFNO models in dataset A, B.}
\centering
{\begin{tabular}{llll} \toprule
Model& Evaluation metric& Dataset A	& Dataset B \\ \midrule
\multirow{2}{2cm}{CNN-GRU} & $rLSE$&34.92&5.68\\
& PCC& 0.4704& 0.9183\\
\midrule
 \multirow{2}{2cm}{BFNO} & $rLSE$& 4.52& 2.04\\
 & PCC& 0.9351& 0.9659\\
  \bottomrule
\end{tabular}}
\label{table3}
\end{table}
\section{Conclusions}
\noindent
In the evaluation of track geometry, considering the vehicle response to track defects requires fast and reliable vehicle-track coupled dynamic simulations based on a large amount of track data. The use of dynamic simulations is well established, but it is computationally expensive. In this paper, the BFNO model was proposed, which provides a more convenient and economical method for estimating vehicle responses and has the potential to be applied to in-service trains. The findings of this study are as follows.
\par
(1)	The proposed BFNO model separates the car body from the bogie and extracts the feature relationship between input and output data in frequency domain. The model demonstrates good estimation performance in both the time and frequency domains. Under the optimal hyperparameter combination, the $rLSE$ of BFNO model is 2.04\%, which is reduced by 64\%, compared with the CNN-GRU model.
\par
(2)	The proposed BFNO model is superior to commercial software in terms of efficiency. In the evaluation of track geometry, users can use pre-trained BFNO to obtain the dynamic response with almost no computational cost. Besides, the improvement in the computational efficiency can support many railway engineering tasks that require repetitive forward numerical simulations.
\par
(3)	Compared with the CNN-GRU model, the BFNO model can make predictions at unseen time steps. The BFNO model is trained on data with a time step of 0.003s and tested on data with a time step of 0.001s, achieving good generalization performance.
\par
However, our proposed model still faces some challenges, such as the prediction performance of the lateral response is slightly inferior to that of the vertical response, the prediction accuracy cannot be guaranteed for physical parameters outside the range of this paper. In the future, we will explore methods to further improve the prediction accuracy of our proposed model and training strategies for model in the case of sparse data. Furthermore, establishing a vehicle-track coupled model under the combined effects of random and disease irregularity, and extend the application of the proposed model in this paper to the analysis and assessment of dynamic responses under defective conditions.
\section*{Declarations}
\noindent
The authors declare that they have no known competing financial interests or personal relationships that could have appeared to influence the work reported in this paper.
\section*{Funding}
\noindent
This study was funded by Key R\&D projects of the Ministry of Science and Technology NO. 2022YFB2602905, Sichuan Nature and Science Foundation Innovation Research Group Project NO. 2023NSFSC1975, and the Fundamental Research Funds for the Central Universities NO. A0920502052301-369.

\begin{appendices}
\section{The range of physical parameters}
\begin{table}[H]
\caption{The range of physical parameters.}
\centering
{\begin{tabular}{lll} \toprule
Notation&Parameter&Range\\ 
\midrule
$M_c$&Car body mass (kg)&[23040, 48000]\\
$M_t$&Bogie mass (kg)&[2080, 3840]\\
$M_w$&Wheelset mass (kg)&[1120, 2400]\\
$I_{cx}$&Car body roll moment of inertia($\mathrm{kg} \cdot \mathrm{m}^{2}$)&[60048, 153965]\\
$I_{cy}$&Car body pitch moment of inertia($\mathrm{kg} \cdot \mathrm{m}^{2}$)&[1128960, 2732400]\\
$I_{cz}$&Car body yaw moment of inertia ($\mathrm{kg} \cdot \mathrm{m}^{2}$)&[1065370, 2732400]\\
$I_{tx}$&Frame roll moment of inertia($\mathrm{kg} \cdot \mathrm{m}^{2}$)&[1684, 3110]\\
$I_{ty}$&Frame pitch moment of inertia($\mathrm{kg} \cdot \mathrm{m}^{2}$)&[1140, 3840]\\
$I_{tz}$&Frame yaw moment of inertia($\mathrm{kg} \cdot \mathrm{m}^{2}$)&[2080, 3840]\\
$I_{wx}$&Wheelset roll moment of inertia($\mathrm{kg} \cdot \mathrm{m}^{2}$)&[576, 1098]\\
$I_{wy}$&Wheelset pitch moment of inertia($\mathrm{kg} \cdot \mathrm{m}^{2}$)&[67.2, 168]\\
$I_{wz}$&Wheelset yaw moment of inertia($\mathrm{kg} \cdot \mathrm{m}^{2}$)&[592, 1176]\\
$K_{px}$&Longitudinal stiffness of primary suspension per axle side (MN/m)&[8, 17.64]\\
$K_{py}$&Lateral stiffness of primary suspension per axle side (MN/m)&[4, 8.97]\\
$K_{pz}$&Vertical stiffness of primary suspension per axle side (MN/m)&[0.44, 2.1264]\\
$K_{sx}$&Longitudinal stiffness of secondary suspension per bogie side (MN/m)&[0.12, 0.2269]\\
$K_{sy}$&Lateral stiffness of secondary suspension per bogie side (MN/m)&[0.12, 0.2269]\\
$K_{sz}$&Vertical stiffness of secondary suspension per bogie side (MN/m)&[0.092, 1.1889]\\
$C_{pz}$&Vertical damping of secondary suspension per truck side($\mathrm{kN} \cdot \mathrm{s} / \mathrm{m}$)	&[4.8, 70.56]\\
$C_{sy}$&Lateral damping of secondary suspension per bogie side($\mathrm{kN} \cdot \mathrm{s} / \mathrm{m}$)	&[7.84, 72]\\
$C_{sz}$&Vertical damping of secondary suspension per bogie side($\mathrm{kN} \cdot \mathrm{s} / \mathrm{m}$)	&[7, 144]\\
  \bottomrule
\end{tabular}}
\end{table}

\section{CNN model architecture}
\noindent
The shape of the input is (10000,3000,40) and the three positions correspond to the number of samples, the length of time and the dimension of the input in Figure~\ref{fig4}. The shape of the remaining layers has the same meaning as above. Lifting denotes the linear transformation to lift the input to the high dimensional space, and the projection back to original space. Conv1d denotes a 1D convolutional layer, BN denotes a batch normalization layer, Elu denotes a rectified linear layer.

\begin{table}[H]
\caption{CNN model architecture.}
\centering
{\begin{tabular}{lll} \toprule
Part&Layer&Output Shape\\ 
\midrule
Input&-&(10000,3000,40)\\
Lifting&Linear&(10000,3000,120)\\
Permute&Permute&(10000,120,3000)\\
Conv 1&Conv1d/BN/Elu&(10000,120,3000)\\
Conv 2&Conv1d/BN/Elu&(10000,120,3000)\\
Conv 3&Conv1d/BN/Elu&(10000,120,3000)\\
Conv 4&Conv1d/BN/Elu&(10000,120,3000)\\
Permute&Permute&(10000,3000,120)\\
Projection 1&Linear/Elu	&(10000,3000,100)\\
Projection 2&Linear/Elu	&(10000,3000,100)\\
Projection 3&Linear	&(10000,3000,15)\\
  \bottomrule
\end{tabular}}
\end{table}

\section{CNN-GRU model architecture}
\noindent
Lifting denotes the linear transformation to lift the input to the high dimensional space, and the projection back to original space. Conv1d denotes a 1D convolutional layer, BN denotes a batch normalization layer, GRU denotes a gated recurrent unit layer, Elu denotes a rectified linear layer.
\begin{table}[H]
\caption{CNN-GRU model architecture.}
\centering
{\begin{tabular}{lll} \toprule
Part&Layer&Output Shape\\ 
\midrule
Input& -	&(10000,3000,40)\\ 
Lifting&	Linear&	(10000,3000,120)\\ 
Permute&	Permute&	(10000,120,3000)\\ 
Conv 1&	Conv1d/BN/Elu&	(10000,120,3000)\\ 
Conv 2&	Conv1d/BN/Elu&	(10000,120,3000)\\ 
Permute&	Permute&	(10000,3000,120)\\ 
GRU 1&	GRU/BN/Elu&	(10000,3000,100)\\ 
GRU 2&	GRU/BN/Elu&	(10000,3000,100)\\ 
Projection 1&	Linear/Elu&	(10000,3000,100)\\ 
Projection 2&	Linear&	(10000,3000,15)\\ 
  \bottomrule
\end{tabular}}
\end{table}

\section{FNO model architecture}
\noindent
Lifting denotes the linear transformation to lift the input to the high dimensional space, and the projection back to original space. Fourier 1d denotes the 1D Fourier operator, Add operation adds the outputs together.
\begin{table}[H]
\caption{FNO model architecture.}
\centering
{\begin{tabular}{lll} \toprule
Part&Layer&Output Shape\\ 
\midrule
Input&	-&	(10000,3000,40)\\ 
Lifting&	Linear&	(10000,3000,120)\\ 
Permute&	Permute&	(10000,120,3000)\\ 
Fourier 1&	Fourier1d/Conv1d/Add/Elu&	(10000,120,3000)\\ 
Fourier 2&	Fourier1d/Conv1d/Add/Elu&	(10000,120,3000)\\ 
Fourier 3&	Fourier1d/Conv1d/Add/Elu&	(10000,120,3000)\\ 
Fourier 4&	Fourier1d/Conv1d/Add/Elu&	(10000,120,3000)\\ 
Permute&	Permute&	(10000,3000,120)\\ 
Projection 1&	Linear/Elu&	(10000,3000,100)\\ 
Projection 2&	Linear/Elu&	(10000,3000,100)\\ 
Projection 3&	Linear&	(10000,3000,15)\\ 
  \bottomrule
\end{tabular}}
\end{table}

\section{BFNO model architecture}
\noindent
The BFNO architecture is divided into the car body branch and the bogie branch, and the final output is combined through an Add layer.

\begin{table}[H]
\caption{BFNO model architecture.}
\centering
{\begin{tabular}{llll} \toprule
Part&	Layer&	Car body Branch Output Shape&	Bogie Branch Output Shape\\ 
\midrule
Input&	-&	(10000,3000,40)&	(10000,3000,40)\\ 
Lifting&	Linear&	(10000,3000,120)&	(10000,3000,120)\\ 
Permute&	Permute&	(10000,120,3000)&	(10000,120,3000)\\ 
Fourier 1&	Fourier1d/Conv1d/Add/Elu&	(10000,120,3000)&	(10000,120,3000)\\ 
Fourier 2&	Fourier1d/Conv1d/Add/Elu&	(10000,120,3000)&	(10000,120,3000)\\ 
Fourier 3&	Fourier1d/Conv1d/Add/Elu&	(10000,120,3000)&	(10000,120,3000)\\ 
Fourier 4&	Fourier1d/Conv1d/Add/Elu&	(10000,120,3000)&	(10000,120,3000)\\ 
Permute&	Permute&	(10000,3000,120)&	(10000,3000,120)\\ 
Projection 1&	Linear/Elu&	(10000,3000,100)&	(10000,3000,100)\\ 
Projection 2&	Linear/Elu&	(10000,3000,100)&	(10000,3000,100)\\ 
Projection 3&	Linear&	(10000,3000,5)&	(10000,3000,10)\\ 
Add&Add&\multicolumn{2}{c}{(10000,3000,15)}\\
  \bottomrule
\end{tabular}}
\end{table}
\end{appendices}

\bibliographystyle{unsrt}  
\bibliography{references}  

\end{document}